\documentclass[12pt]{iopart}
\usepackage{amsfonts}
\usepackage{amssymb}
\usepackage{graphicx}

\usepackage{indentfirst}

\begin{document}
\title{Strong field double ionization: What is under the ``knee''?}

\author{F. Mauger$^{1,2}$, C. Chandre$^2$, T. Uzer$^3$}
\address{$^1$ Ecole Centrale de Marseille, Technop\^ole de Ch\^ateau-Gombert, 38 rue Fr\'ed\'eric Joliot Curie
F-13451 Marseille Cedex 20, France
\\$^2$ Centre de Physique Th\'eorique, CNRS -- Aix-Marseille Universit\'es, Campus de Luminy, case 907, F-13288 Marseille cedex 09, France \\ $^3$ School of Physics, Georgia Institute of Technology, Atlanta, GA 30332-0430, USA}

\begin{abstract}
Both uncorrelated (``sequential'') and correlated (``nonsequential'') processes contribute to the double ionization of the helium atom in strong laser pulses. The double ionization probability has a characteristic ``knee'' shape as a function of the intensity of the pulse. We investigate the phase-space dynamics of this system, specifically by finding the dynamical structures that regulate the ionization processes. The emerging picture complements the recollision scenario by clarifying the distinct roles played by the recolliding and core electrons. Our analysis leads to verifiable predictions of the intensities where qualitiative changes in ionization occur, leading to the hallmark ``knee'' shape.
\end{abstract}
\submitto{\JPB}
\maketitle

\section{Introduction}

\begin{figure}
\center
        \includegraphics[width=80mm]{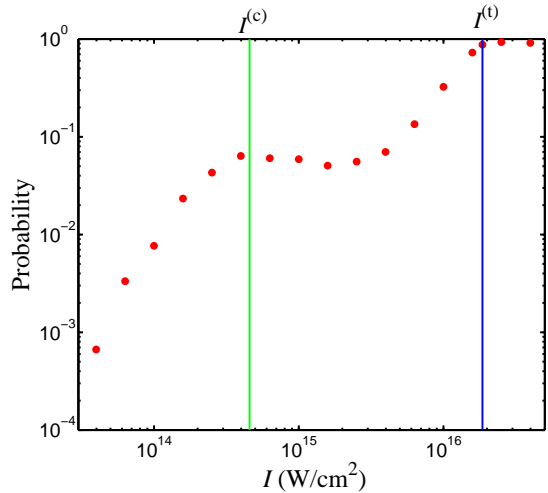}
        \caption {\label{fig:1}   
        Double ionization probability for Hamiltonian~(\ref{Hamiltonian}) for $\omega=0.0584$ as a function of the intensity of the field $I$. We use a microcanonical initial set of~$3000$ trajectories randomly chosen in the accessible phase space of the helium atom in the ground state of energy. The field is a 2-4-2 laser cycle pulse shape (see Fig.~\ref{fig:3}).
        The vertical lines indicate (in green) the laser intensity $I^{(c)}\approx 4.58 \times 10^{14}\ \mbox{W} \cdot \mbox{cm}^{-2}$ where our dynamical analysis predicts the maximum of nonsequential double ionization, and (in blue) the intensity $I^{(t)}\approx 1.86 \times 10^{16}\ \mbox{W} \cdot \mbox{cm}^{-2}$ where the double ionization is expected to be complete.}   
\end{figure}

Atoms and molecules subjected to strong laser pulses generically display multiple electron ionizations. The three ionization processes for the simplest system with two electrons, the helium atom, are~: \\
Single ionization (SI)~:
$$
\mbox{He} \rightarrow \mbox{He}^{+} + e^{-},
$$
Sequential double ionization (SDI)~:
$$
\mbox{He} \rightarrow \mbox{He}^{+} + e^{-} \rightarrow \mbox{He}^{2+} + 2e^{-},
$$
Nonsequential double ionization (NSDI)~:
$$
\mbox{He} \rightarrow \mbox{He}^{2+} + 2e^{-}. 
$$
Surprisingly, at some intensities correlated (nonsequential) double ionization rates are 
several orders of magnitude higher than the uncorrelated 
sequential mechanism leads one to believe, making the characteristic ``knee'' shape in the 
double ionization yield versus intensity plot into one of the 
most dramatic manifestations of electron-electron correlation in nature. 
Different scenarios 
have been proposed to explain the mechanism behind this surprise~\cite{fitti92,cork93,scha93,walk94,beck96,kopo00,lein00,sach01,fu01,panf01,barn03,colg04,ho05_1,ho05_2,ruiz05,horn07,prau07,feis08}. When confronted with experiments~\cite{brya06,webe00_2}, 
the recollision scenario~\cite{cork93,scha93}, in which an ionized electron is hurled back at the core and ionizes the second electron, seems in best accord with observations. Numerous articles study the characteristic behavior of the atom-field interaction through quantum~\cite{lein00,wats97}, semi-classical~\cite{fu01,chen03,brab96} or classical~\cite{panf02,ho05_3,ye08,haan08,sach01} mechanics.
In Fig.~\ref{fig:1}, a generic double ionization probability as a 
function of the intensity of the laser field is plotted based on the work in this article.
Similar knees have been observed in 
experimental data~\cite{fitti92,kodo93,walk94,laro98,webe00_2,corn00,guo01,dewi01,ruda04} and successfully 
reproduced by quantal computations on atoms and molecules~\cite{beck96,wats97,lapp98,panf03}.
In a recent series of articles ~\cite{fu01,sach01,panf02,panf03,ho05_1,ho05_2,liu07} characteristic 
features of double ionization were reproduced using classical trajectories
and this success was ascribed to the dominant role of correlation~\cite{ho05_1}. Indeed, entirely classical interactions turn out to be adequate to generate the strong two-electron correlation needed for double ionization. 

In this article, we complement the recollision scenario by answering two crucial open questions: "What does the core electron do during the recollision process?" and "How do the two electron share the energy brought back by the recolliding electron?". Our global view of the dynamics based on modern tools of nonlinear dynamics leads to two verifiable predictions for key points which make up the knee  
in Fig.~\ref{fig:1}: The laser intensity where nonsequential double ionization is maximal and the intensity where the double ionization is complete~\cite{beck96,lapp98}. These results were recently announced in a Letter~\cite{maug09}.

In Sec.~\ref{sec:class}, we give some details on the classical one-dimensional Hamiltonian model we use for our study. In Sec.~\ref{sec:Hhe}, we analyze the dynamics of this model without the field, by first considering the uncorrelated motion and then the correlated motion. This recognition leads to the definition of an ``inner'' and an ``outer'' electron. In Sec.~\ref{sec:field}, the dynamics of these inner and outer electrons is analyzed with the laser field. By identifying reduced models and their organizing structures for the three different ionization processes (SI, SDI, NSDI), the characteristic features of the ionization probability are obtained.

\section{Classical model of the helium atom}
\label{sec:class}

\subsection{Hamiltonian model with a soft potential}

We consider here a fully classical one-dimensional model for the helium atom. It has two degrees of freedom, each of one associated with the two distances between the nucleus and the two electrons. The Hamiltonian is composed of a kinetic energy plus three soft Coulomb potentials associated with the three pairs of charged particles (the so-called Rochester potential~\cite{rochester1,rochester2}), where the soft Coulomb potential is used to remove the singularities~\cite{ho05_1}. The Hamiltonian is, in atomic units~(a.u.)~:
\begin{eqnarray}
  && {\mathcal H}_e(x,y,p_{x}, p_{y}) =  \frac{p_{x}^{2}}{2} + \frac{ p_{y}^{2}}{2} \nonumber \\
                          && \quad  +\frac{1}{\sqrt{(x-y)^{2}+\alpha}}
                             -\frac{2}{\sqrt{x^{2}+\alpha}} -\frac{2}{\sqrt{y^{2}+\alpha}},
                             \label{Hamiltonian_he}
\end{eqnarray}
where $x$, $y$ and $p_{x}$, $p_{y}$ are respectively the positions and (canonically conjugate) momenta of each electron.
We assume that the soft parameter $\alpha$ is equal to one (by appropriate rescalings or according to Refs.~\cite{ho05_1,panf01,panf03,panf02,ho05_3}). In what follows, we fix the energy at the ground state energy ${\mathcal H}_e={\mathcal E}_g=-2.24$~a.u.~\cite{ho05_1}.

The dynamics of Hamiltonian ${\mathcal H}_e$ is investigated in Sec.~\ref{sec:Hhe}. It is shown to be very chaotic with a few hyperbolic periodic orbits organizing its dynamics~\cite{chaosbook}. This will lead to the definition of an inner and an outer electron. Here, since its phase space is bounded at the ground state energy, no ionization occurs.

\begin{figure}
\center
        \includegraphics[width=80mm]{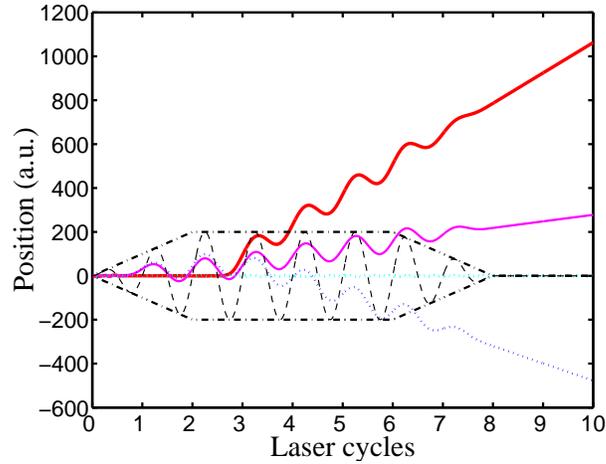}
        \caption{\label{fig:3}       
        Example of double (full line curves, red online) and single (dotted lines, blue online) ionization trajectories of Hamiltonian~(\ref{Hamiltonian}) for $\omega = 0.0584$ and $I = 10^{15} \ \mbox{W} \cdot \mbox{cm}^{-2}$. The position of each pair of electrons is plotted versus time. The pulse shape function~$f(t)$ (dotted dashed black curve) and the laser excitation~$E(t)$ (dashed black curve) are also represented (dashed-dotted and dashed curves). The amplitude of the shape function and the laser excitation are not representative of the actual conditions.}
\end{figure}

Secondly, we consider an external laser field driving the helium atom. The interaction with the laser field is modeled through the function~$E(t)$ and the Hamiltonian is given by~\cite{panf01}:
\begin{eqnarray}
    && {\mathcal H}(x,y,p_{x}, p_{y},t) =  \frac{p_{x}^{2}}{2} + \frac{ p_{y}^{2}}{2}+(x+y)E(t) \nonumber \\
    &&  \quad                       +\frac{1}{\sqrt{(x-y)^{2}+1}}
                             -\frac{2}{\sqrt{x^{2}+1}} -\frac{2}{\sqrt{y^{2}+1}}.
                             \label{Hamiltonian}
\end{eqnarray}
As a laser field, we consider a sinusoidal pulse with an envelope, i.e.\ $E(t)= E_{0} \ f(t) \ \sin \omega t$ where $E_{0}$ is the peak field strength, $\omega$ the laser frequency and $f(t)$ the pulse shape function. We choose for $f(t)$ a trapezoidal function with 2-4-2 laser pulse shape~\cite{ho05_1,ho05_2,panf03,panf02} as shown on Fig.~\ref{fig:3} (the ramp-up lasts two cycles, the plateau four, and the ramp-down two). In what follows, we choose $\omega=0.0584$~a.u.~(unless specified) which corresponds to a wavelength of 780~nm. The relation between $E_0$ and the intensity of the field is given by $E_0\mbox{(a.u.)}=5.329\cdot 10^{-9} \sqrt{I(\mbox{W} \cdot \mbox{cm}^{-2})}$. Hamiltonian~(\ref{Hamiltonian}) has two and a half degrees of freedom (two degrees of freedom coming from the helium atom without the field, and the other half degree of freedom coming from the time-dependence of the Hamiltonian introduced by the field).

The purpose of Hamiltonian~(\ref{Hamiltonian}) is to reproduce the experimental data on ionization probability, and more specifically concerning the double ionization~\cite{ho05_1,ho05_2,panf03,panf02,ho05_3}, using classical nonlinear dynamics. The numerical integration of trajectories shows that electrons can be unbounded in presence of the field. These trajectories correspond to ionizing trajectories. Examples of ionized trajectories of Hamiltonian~(\ref{Hamiltonian}) are represented in Fig.~\ref{fig:3}~: The dashed curves (blue and cyan) refer to a single ionization. The continuous curves (red and magenta) correspond to a nonsequential double ionization.

\begin{figure}
\center
        \includegraphics[width=80mm]{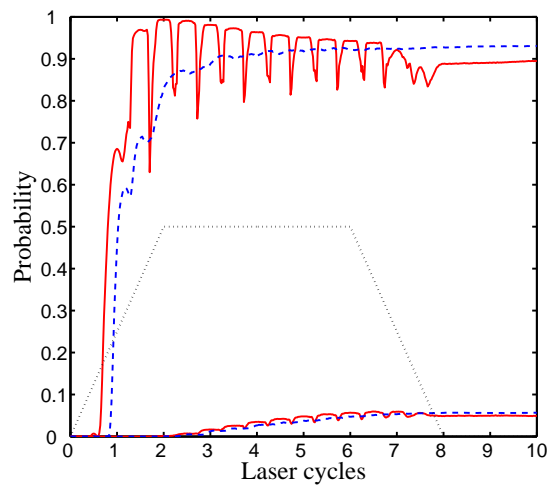}
        \caption{\label{fig:2}
        Probability of simple (upper curves) and double (lower curves) ionization versus time for Hamiltonian~(\ref{Hamiltonian}). The frequency of the pulse is $\omega = 0.0584$~a.u. and the intensity is~$I = 10^{15} \ \mbox{W} \cdot \mbox{cm}^{-2}$. Two different criteria for ionization are compared~: a criterion based on the energy~\cite{panf02} (continuous lines, red online), and a criterion based on the distance of both electrons from the nucleus (dashed lines, blue online). The pulse shape function $f(t)$ is plotted with arbitrary units (dotted line).}
\end{figure}

There are two qualitatively different ways to obtain double ionization~: Nonsequential double ionization, where the two electrons leave the core (inner) region at about the same time, and sequential double ionization, where one electron leaves the inner region long time after the other one has ionized. 
In order to compute the ionization probability, we use a distance criterion for ionization~: When an electron is further than $30$~a.u.\ from the nucleus, it is said to be ionized. 
Using the distance criterion, single ionization corresponds to the case where one electron coordinate, e.g. $x$, is larger than~$30$~a.u.\ while $y$ is smaller than~$30$~a.u. Double ionization corresponds to the case when both $x$ and $y$ are larger than~$30$~a.u.
Other choices of ionization criteria lead to qualitatively (and even quantitatively) similar results. In Fig.~\ref{fig:2} we show the single and double ionization probabilities as functions of time for $I=10^{15}\ \mbox{W} \cdot \mbox{cm}^{-2}$ and $\omega=0.0584$ using two different definitions for ionized trajectories. In full and dashed curves (red online) we consider a criterion based on the energy~\cite{panf02}. In dotted and dashed dotted curves (blue online) we consider the above ionization criterion based on the distance. We see clearly that the two sets of curves display qualitatively the same information. The time at which most ionization occurs depends slightly on this definition. The delay between the two curves is due to the time it takes to sweep away the two electrons, once they have acquired enough energy. While the single ionization is triggered at a determined time where its probability abruptly becomes predominant, the double ionization is more regular and increases after about two laser cycles (which is consistent with quantum simulations~\cite{dund99}). 

\subsection{Initial conditions}

We first consider Hamiltonian~(\ref{Hamiltonian}) without the laser field ($E_0=0$).
A typical trajectory reveals a highly chaotic behavior, filling very quickly the whole accessible region in phase space. This observation would lead one to characterize the system as very close to ergodic. This dynamical property enables one to compile statistics on ionization with different kinds of initial sets. For instance, in Ref.~\cite{ho05_1,panf01,ho05_2,ho05_3}, the authors first integrate one trajectory without the field, and then consider the points of this trajectory to generate their initial set. The second choice of initial set is a microcanonical initial distribution which is obtained by choosing randomly points over the accessible phase space. Since there is no apparent regular (stable) structures (see the Finite-Time Lyapunov maps in Fig.~\ref{fig:5} below) the two kinds of initial sets lead to approximately similar results if the integration time as well as the number of points are sufficiently large. Due to this chaotic property, we can also select partial microcanonical distributions (randomly chosen points of a small portion of the accessible region of phase space) as other sets of initial conditions. We have checked that all these initial sets lead to the same ionization probability curve shown in Fig.~\ref{fig:1}. We notice, though, that for the microcanonical distributions, the ionization probability curves converge faster (with typically 3000 trajectories) than the ones associated with the distributions from the free motion generation (which require 10000 trajectories to reproduce similar smooth results).
In this article, we use a microcanonical initial distribution over the entire admissible phase space whenever statistical indicators are concerned.

\section{Classical mechanisms of single and double ionization}
\label{sec:resu}

From the numerical integration of a large assembly of particles, statistical indicators can be computed, ionization probabilities as it was done for Fig.~\ref{fig:1} are an example of such statistical indicators. The main drawback of statistical analysis is that it provides very little information on the dynamics and hence on the physical mechanisms.
 Other methods use not only the locations of trajectories but also their (linear and nonlinear) stability properties. More sophisticated methods find the organizing centers of the dynamics which are invariant structures such as periodic orbits, invariant tori, stable and unstable manifolds of such objects. In what follows, we apply these tools of dynamical systems theory to the helium atom in order to infer the classical mechanisms of single and double ionizations.

\subsection{Helium atom without external field} \label{sec:Hhe}

First we study at the dynamics of the helium atom without external field. This analysis is performed to describe the possible states of the system when the laser field is turned on. Two conclusions can be drawn from this analysis~: First, that the dynamics without the field is very chaotic and second, that the organizing centers are four hyperbolic periodic orbits. These findings lead to the definition of an inner and an outer electron once the laser field is turned on.  

\subsubsection{Uncorrelated motion}

\begin{figure}
        \includegraphics[width=80mm]{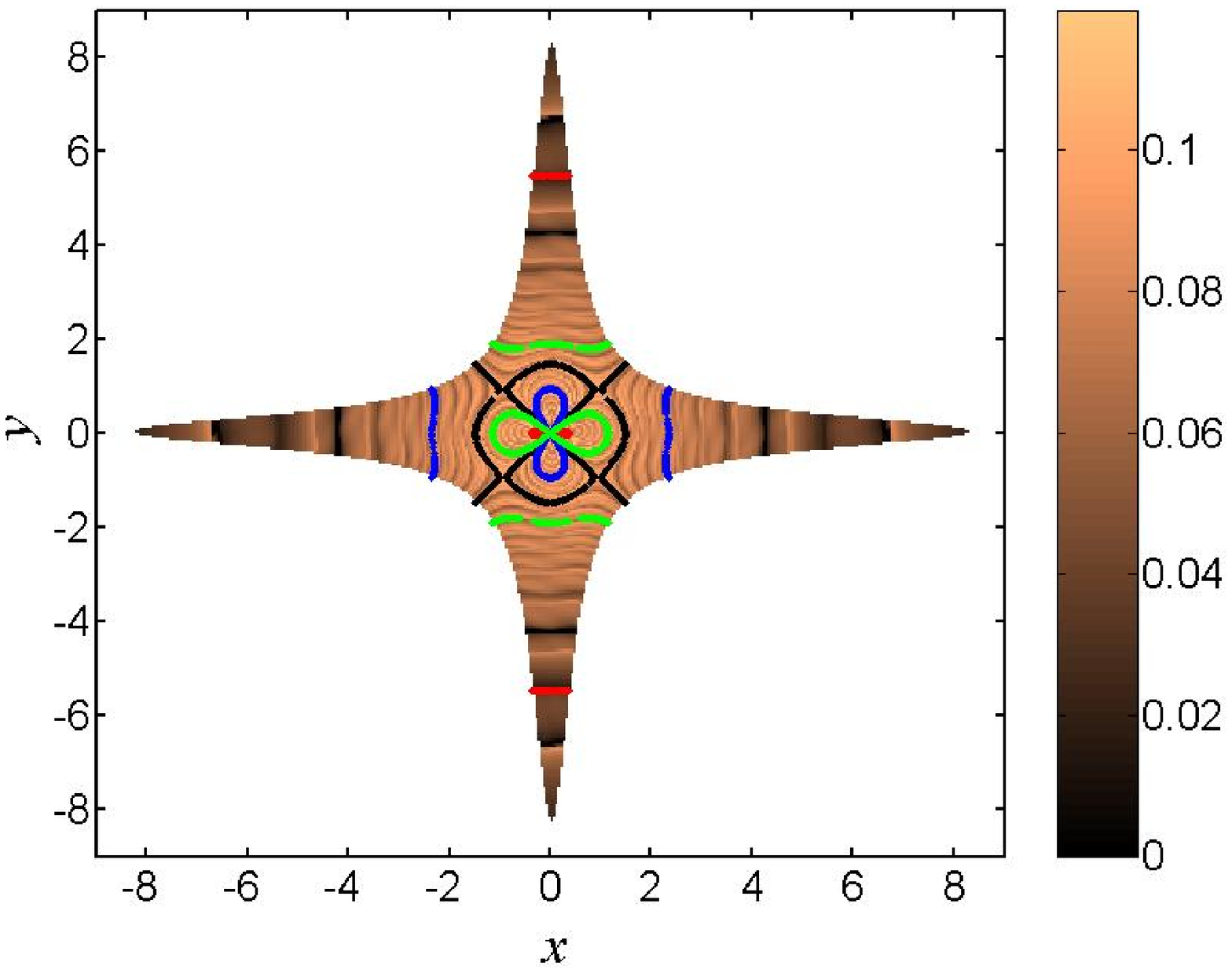}
        \includegraphics[width=80mm]{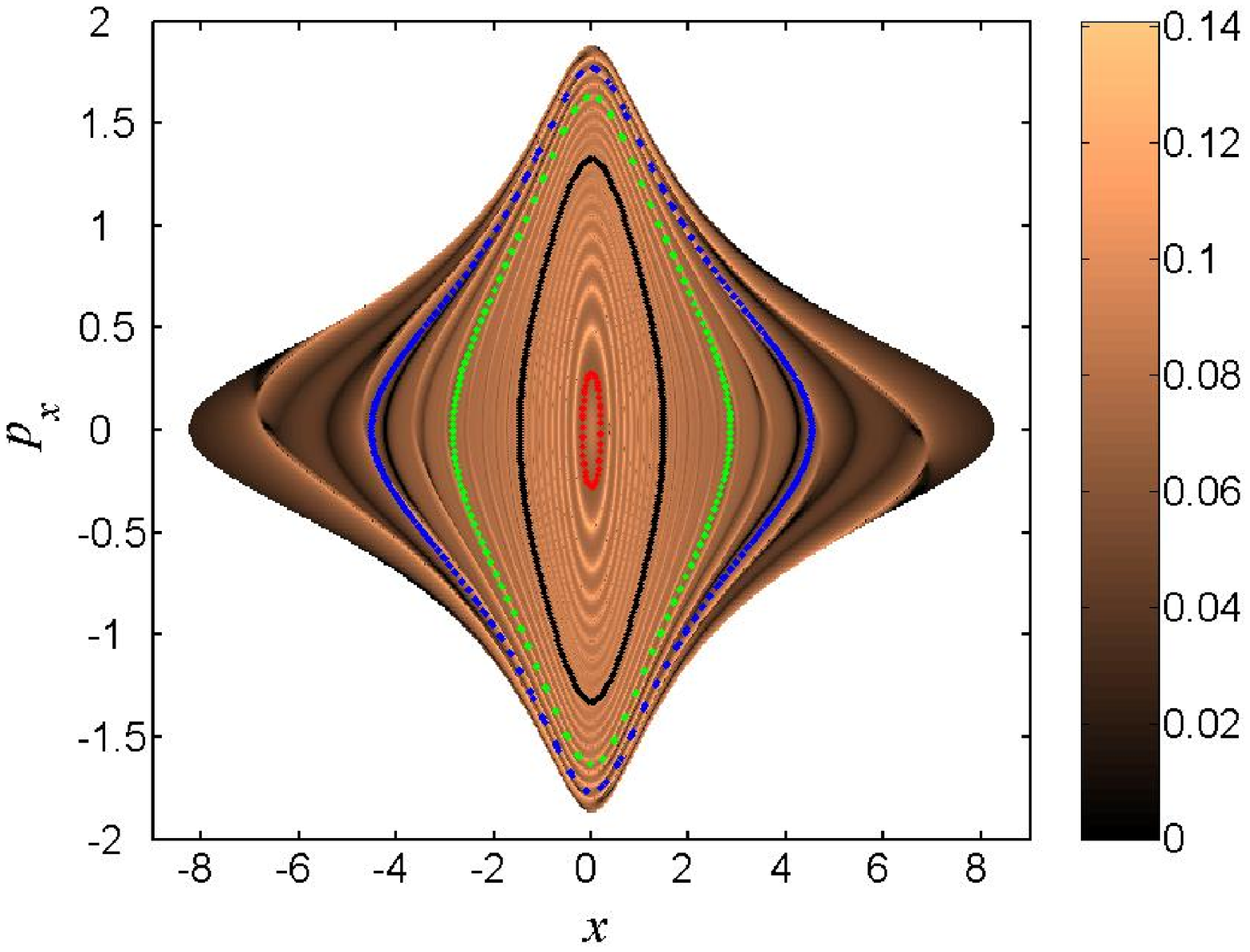}
        \caption{\label{fig:4}
        FTL map of Hamiltonian~(\ref{Hamiltonian_he}) without the electron-electron interaction term at time $t=43$ a.u.~in the $(x,y)$ (left panel) and $(x,p_{x})$ (right panel) planes. We also plot the Poincar\'{e} sections of some trajectories (continuous lines)~: On the left panel, the equation of the section is $x p_{x} + y p_{y} =0$. On the right panel, it is~$y=0$.
        }
\end{figure}

Without the electron-electron interaction term $1/\sqrt{(x-y)^{2}+1}$ in Hamiltonian~(\ref{Hamiltonian_he}), the resulting Hamiltonian is integrable since it is the sum of two independent systems, each with one degree of freedom. The motion occurs on products of two periodic orbits, each of them associated with one electron.
Since this model helium atom (without external field) has two degrees of freedom, it is natural to consider Poincar\'e sections of trajectories, two of which are shown in Fig.~\ref{fig:4}. In the left panel, the section in the plane $(x,y)$ has equation $xp_x+yp_y=0$, and in the right panel, the section in the plane $(x,p_x)$ of one single electron has the equation $y=0$. 
Another representation of phase space is afforded by the linear stability analysis of the trajectories as given by the Finite-Time Lyapunov (FTL) exponents~\cite{froe97,guzz02,okus03,chaosbook}.
The FTL exponents are obtained by integrating the tangent flow together with the equations of motion for ${\bf X}=(x,y,p_x,p_y)$~:
\begin{eqnarray}
   \dot{\bf X} & = & {\bf F}({\bf X},t),          \label{Eq_motion} \\
   \dot{J} & = & DF({\bf X},t)  J, \label{Tangent_flow}
\end{eqnarray}
where Eq.~(\ref{Eq_motion}) are the equations of motion, and Eq.~(\ref{Tangent_flow}) is the tangent flow where $DF({\bf X},t)$ is the matrix of variations of the generalized velocity field ${\bf F}$ at the point~$\bf X$ and time $t$, i.e.\ $DF_{ij}=\partial F_i/\partial X_j$. The initial condition for the integration of the tangent flow is $J_{0} = \mathbb{I}_{4}$, the four dimensional identity matrix. The (maximum) FTL exponent at time $t$ for the initial conditions ${\bf X}_0$ is equal to $l(t;{\bf X}_0) = \log|\lambda(t;{\bf X}_0)|/t$ where~$\lambda(t;{\bf X}_0)$ is the eigenvalue of the Jacobian matrix $J$ at time~$t$ with the largest norm. The way to analyze the dynamics using these exponents is to represent maps of FTL exponents as functions of the initial conditions ${\bf X}_0$ at a fixed time $t$. These maps (called FTL maps) quantify the (linear) instability of some regions and highlight invariant objects as it is shown in Fig.~\ref{fig:4}. In this respect, they display a dynamical information which complements Poincar\'e sections which would highlight the invariant structures and the size of the chaotic zones in phase space.

We consider two types of FTL maps, each associated with a different choice of set of initial conditions, and corresponding to the two Poincar\'e sections mentioned above. The first is in the physical space, i.e.\ the $(x,y)$ plane, where we take initial conditions over the surface of equation~$x p_{x} + y p_{y} = 0$, with~$p_{x}$ and~$y$ bearing the same sign. For that, we write~$p_{x}=\beta y$ and~$p_{y}=-\beta x$, and we adapt the parameter~$\beta$ so that the trajectory belongs to the ground state ($\beta \geq 0$). The second set of initial conditions is on the reduced phase space of one electron, e.g., on the $(x,p_{x})$ plane, where we take the initial conditions over the surface of equation~$y=0$ with~$p_{y} \geq 0$. For that, we first choose the initial position~$(x,p_{x})$, and then we adapt the momentum~$p_{y}$ to fulfill the condition on the energy.

We have plotted on Fig.~\ref{fig:4} the FTL maps as well as the Poincar\'e sections of a few trajectories of Hamiltonian~(\ref{Hamiltonian_he}) without the correlation term. We see that the Poincar\'{e} sections match the invariant structures revealed by the FTL maps. In this way, in the $(x,y)$ plane, the Poincar\'{e} section of equation~$x p_{x} + y p_{y} =0$ reproduces the two central ``eight''-shape and the fine structures on the branches of the star (which represents the accessible phase space region). In the $(x,p_{x})$ plane, the section of equation~$y=0$ displays more clearly the dynamical information and also matches the structures of the FTL maps, from the rings in the center, up to the distorted ones on the sides.

This correspondence between Poincar\'{e} sections and the FTL maps means that the FTL maps also identify the invariant structures in phase space as one-dimensional curves with an approximately constant FTL exponent (which is about 0.25). In addition, as we will see below, it reveals linear stability properties and extent of chaos in the system, making these maps an ideal tool and one of our methods of choice to analyze the dynamics.

\subsubsection{Correlated motion}

\begin{figure}
        \includegraphics[width=80mm]{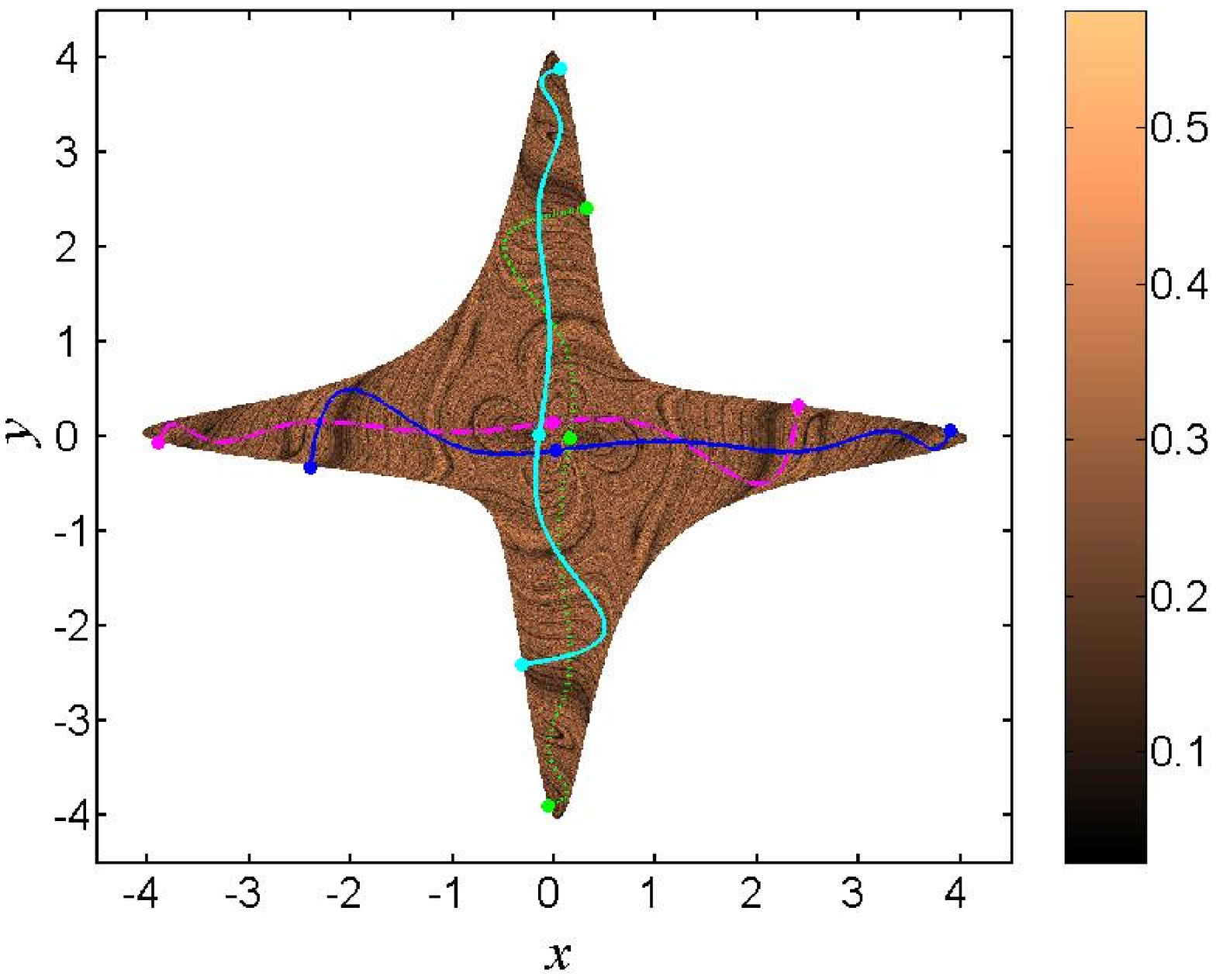}
        \includegraphics[width=80mm]{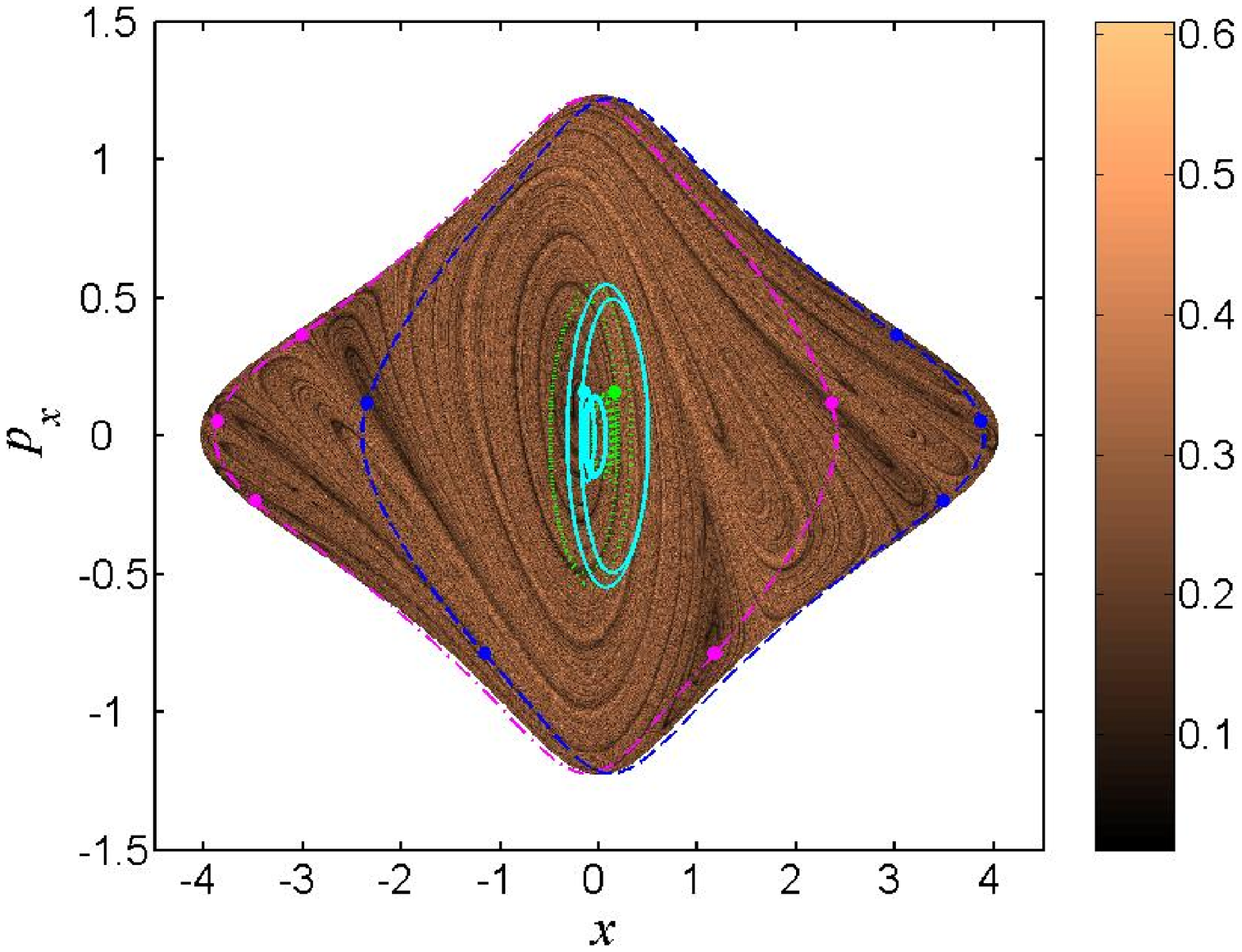}
        \caption{\label{fig:5}
        FTL maps of Hamiltonian~(\ref{Hamiltonian_he}) at time $t=43$ a.u.~in the $(x,y)$ (left panel) and $(x,p_{x})$ (right panel) planes using the same two ensembles of initial conditions as in Fig.~\ref{fig:4}. We also plot the projections of the periodic orbits ~$O_{x,1}$, $O_{x,2}$, $O_{y,1}$ and~$O_{y,2}$ (as continuous curves) and their respective Poincar\'e sections (as dots)~:
$O_{x,1}$ (dashed dotted line, pink online),
$O_{x,2}$ (dashed line, blue online),
$O_{y,1}$ (dotted line, green online),
$O_{y,2}$ (full line, cyan online).
        }
\end{figure}

The addition of the electron-electron interaction term breaks the integrability of the uncorrelated motion. We expect some tori to be broken and some others to be preserved as is typical from Hamiltonian dynamics with bounded phase space. However, the scenario is different. In Fig.~\ref{fig:5} we have represented the FTL maps in both $(x,y)$ and $(x,p_x)$ planes. It appears that the $(x,y)$ FTL map reveals structures very similar to those of the FTL map of the uncorrelated motion. In particular, it reproduces the eight-shape figure at the center and the stripes on the branches of the star. However, we notice that the FTL exponents are relatively high (up to 0.6), at least significantly larger than the uncorrelated motion, indicating a strong chaos. From this map, it is difficult to infer any globally chaotic property or the existence of stable invariant structures. On the contrary, the $(x,p_x)$ FTL map clearly displays a globally chaotic property by showing fine details of stretching and folding of trajectories which is characteristic of a chaotic behavior. In particular, this figure does not display any regular elliptic island of stability as it was the case for the uncorrelated motion. In other words, since the dynamics is very chaotic, Poincar\'e sections do not provide any useful information. At first inspection, these sections do not show any regular motion (like elliptic islands). Each trajectory intersects the Poincar\'e section as scatters of points.

From the inspection of a wide ensemble of trajectories, we notice that the motion is mainly driven by four periodic orbits. The organizing centers of the dynamics are numerically determined using standard periodic orbit search methods (see Ref.~\cite{chaosbook}). The key stept is to select an appropriate Poincar\'e section. Since the flow reduces to a map on this section, the search for periodic orbits reduces to finding zeros of a function in a certain space. Using a Newton-Raphson algorithm, a good initial guess based on a quick inspection of trajectories in phase space converges sufficiently fast to a true periodic orbit of the flow. In order to obtain the linear stability properties, we integrate the tangent flow as described in the section on the FTL maps.

The four important periodic orbits are denoted~$O_{x,1}$, $O_{x,2}$, $O_{y,1}$ and~$O_{y,2}$, and their projections are displayed in Fig.~\ref{fig:5}. These four orbits look alike, and they can be obtained from one single periodic orbit through the symmetries of the equations of motion:~$(x,y,p_{x},p_{y}) \mapsto (y,x,p_{y},p_{x})$, $(x,y,p_{x},p_{y}) \mapsto (-x,-y,-p_{x},-p_{y})$ and~$(x,y,p_{x},p_{y}) \mapsto (-y,-x,-p_{y},-p_{x})$. It also means that the representation of these orbits on the $(y,p_{y})$ plane can be deduced from the one on the $(x,p_{x})$ plane by inverting the coordinates~$x$ and~$y$.

\begin{figure}
\center
        \includegraphics[width=80mm]{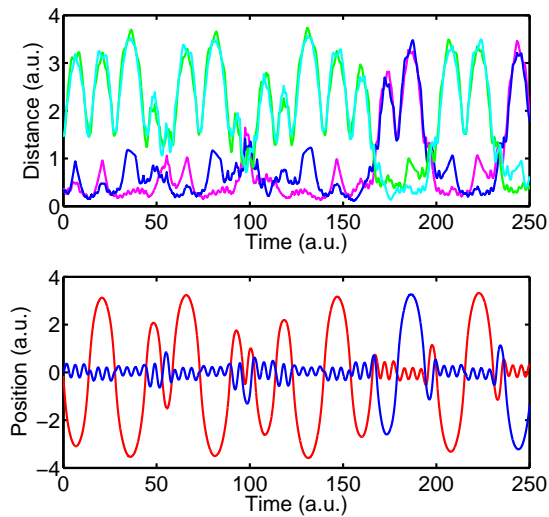}
        \caption{\label{fig:6}
        Upper panel: Distance in phase space versus time, of a typical trajectory of Hamiltonian~(\ref{Hamiltonian_he}) to the four periodic orbits~$O_{x,1}$, $O_{x,2}$, $O_{y,1}$, $O_{y,2}$. For each distance, the color code and line style follow the one in Fig.~\ref{fig:5} (see text). Lower panel: Positions $x$ (red online) and $y$ (blue online) of the two electrons as a function of time for the same trajectory as the upper panel. 
        }
\end{figure}

First we notice that the motion in each direction ($x$ and~$y$) is driven by two periodic orbits (respectively~$O_{x}$ and~$O_{y}$)~: In Fig.~\ref{fig:6} the distance in the phase space of a typical trajectory to the four periodic orbits displayed in Fig.~\ref{fig:5} is represented versus time. We clearly see that the trajectory follows one periodic orbit before following another one since one of the four distances is small most of the time. The corresponding trajectory is represented in the lower panel of Fig.~\ref{fig:6}. We notice that at each time the trajectory changes the periodic orbit it follows, it is associated with a switch of the role between inner and outer electron (for instance, from $t=167$ to $t=194$). 

As usual, the phase space of Hamiltonian~(\ref{Hamiltonian}) is filled with other (likely all hyperbolic) periodic orbits. However, they seem to play a less important role, at least for the short times we consider here. The motivation for studying the impact of these relatively short periodic orbits is that the laser field will only act on the system for a short time (typically of order of 800 a.u.).
These four periodic orbits all have the same period of 29~a.u., therefore much shorter than the duration of the laser pulse whose influence on the dynamics is investigated in the next section. The reasons why they are so important for the dynamics are two-fold~: They are sufficiently short so that a typical trajectory has time to follow it several times during a laser pulse, and they are weakly hyperbolic so that this typical trajectory mimics the dynamics on this orbit if it passes by closely. 

The two outer periodic orbits~$O_{x,1}$ and $O_{x,2}$ in the plane~$(x,p_x)$ have projections on $(y,p_y)$ which are the same as the inner periodic orbits $O_{y,1}$ and $O_{y,2}$ by symmetry. This means that the motion on each of these periodic orbits, and consequently of a typical trajectory is composed of one electron close to the nucleus and the other one further, with quick exchanges of the roles of each electron. This defines at each time an inner electron and an outer one. We will see in the next section that this distinction is crucial since the field will only act on the outer electron to drive ionization.

\subsection{Helium atom driven by an external field}
\label{sec:field}

\subsubsection{Finite-Time Lyapunov and electron ionization maps}

\begin{figure}
\center
        \includegraphics[width=80mm]{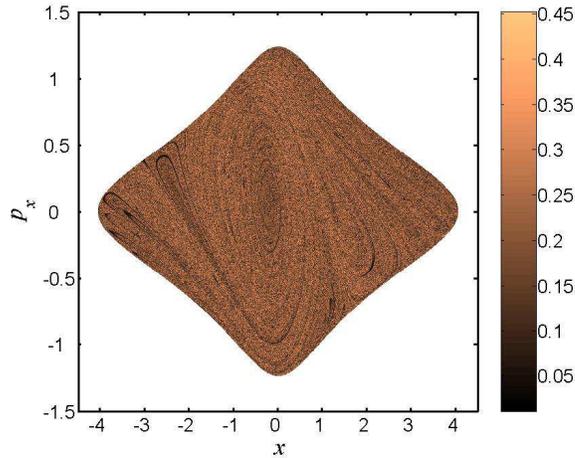}
        \caption{\label{fig:7} 
        FTL map of Hamiltonian~(\ref{Hamiltonian}) for $I=10^{15}\ \mbox{W} \cdot \mbox{cm}^{-2}$ and $\omega=0.0584$ at time $t=107.6$ a.u.\ (one laser cycle).
        }
\end{figure}

We consider the same initial conditions as the ones considered without the laser field in the previous section (see Fig.~\ref{fig:5}). In Fig.~\ref{fig:7}, the FTL map is plotted using the set of initial conditions belonging to the reduced phase space of one electron in the admissible region of the ground state. We notice that this figure is very similar to Fig.~\ref{fig:5} in the sense that it shows the same structures. The dynamics in the presence of the field is also very chaotic, showing a strong dependence on the initial conditions throughout phase space.

\begin{figure}
        \includegraphics[width=80mm]{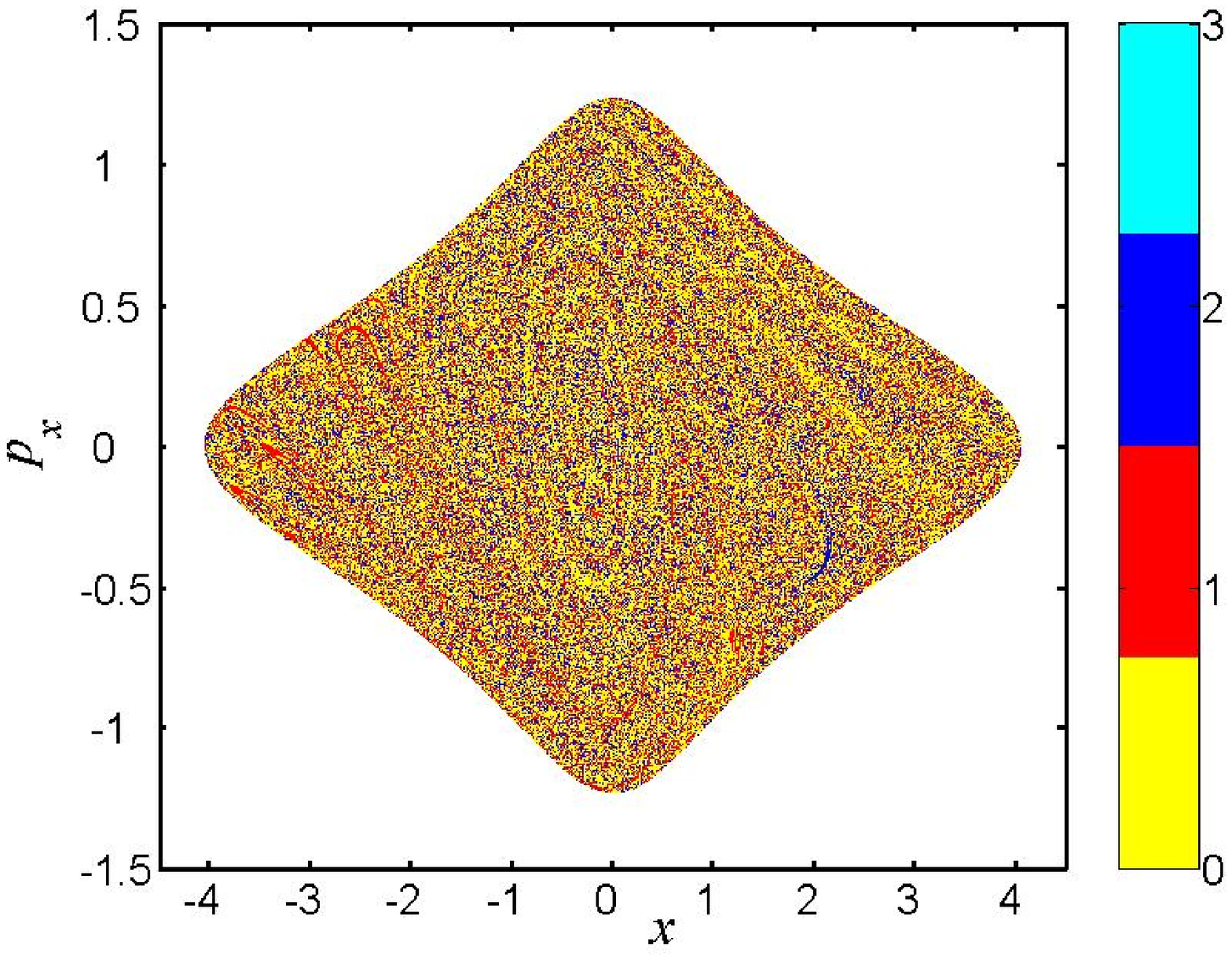}
        \includegraphics[width=80mm]{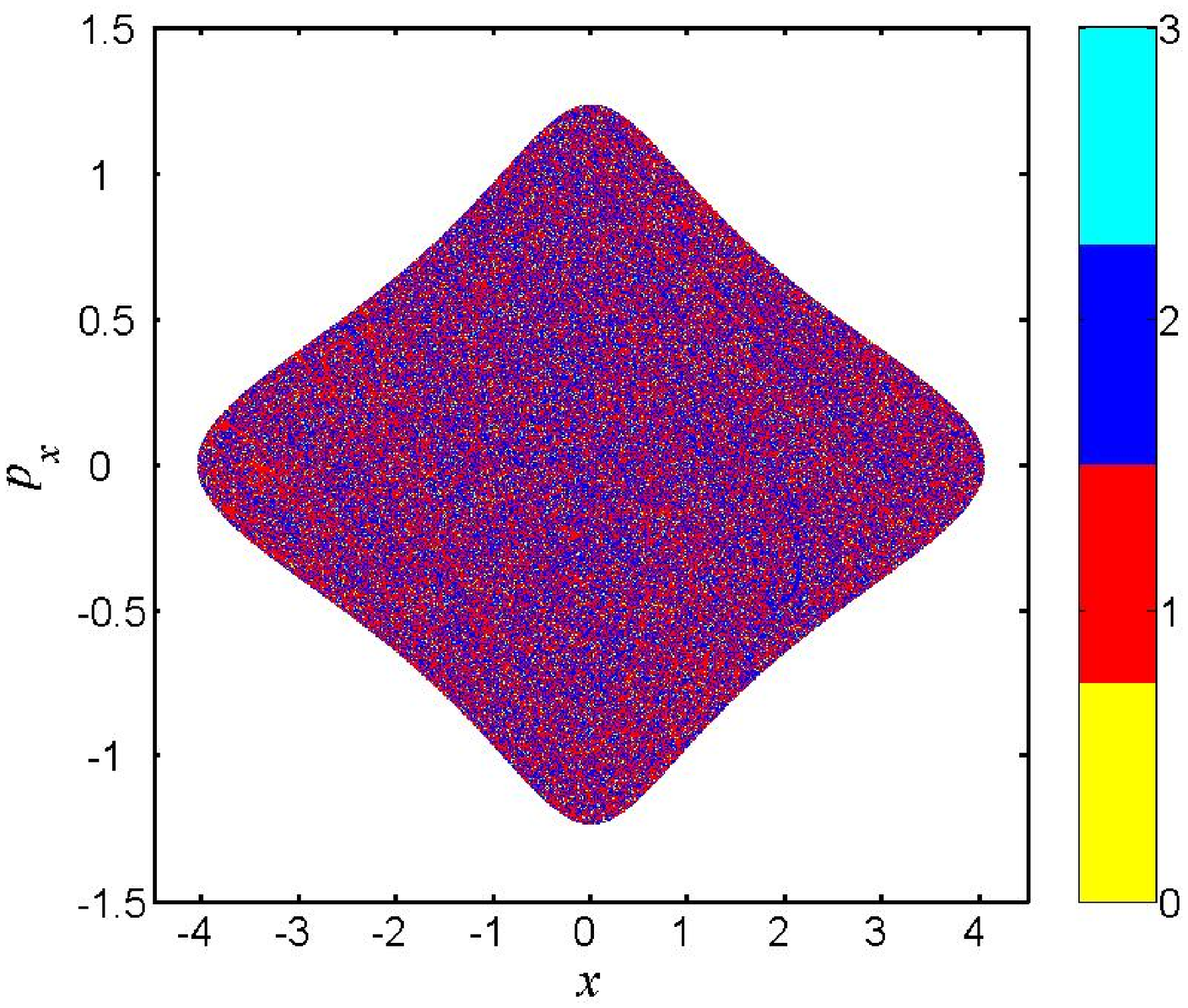}
        \caption{\label{fig:8} 
        Maps of ionized trajectories of Hamiltonian~(\ref{Hamiltonian}) for $I=10^{15}\ \mbox{W} \cdot \mbox{cm}^{-2}$ and $\omega=0.0584$ at times $t=107.6$ a.u.\ (1 laser cycle, left panel) and $t=538$ a.u.\ (5 laser cycles, right panel). The color code is the following: 1 corresponds to the electron labeled by $x$ which has ionized, 2 to the one labeled by $y$ which has ionized, and 3 to the double ionization.
        }
\end{figure}

A natural question is to identify which of these initial conditions lead to ionization, whether it is a single or double ionization. In order to locate these initial conditions, we have plotted in Fig.~\ref{fig:8} the set of initial conditions which lead to ionization after one laser cycle (during the ramp-up of the field) and after five laser cycles (in the plateau of the laser field), using the same set of initial conditions. As expected, after one cycle, there is a significant number of single ionization but no double ionization (upper panel). This is consistent with Fig.~\ref{fig:2}. Double ionization occurs in the middle of the plateau of the laser field (lower panel). We notice that these plots for single ionization reproduce the structures observed on the FTL maps without the field (since this ionization appears early, it is natural to expect some remnants of the phase space structures of the helium atom). However, double ionization occurs apparently uniformly in the set of initial conditions without showing any structure. The structures observed earlier for the helium atom have been washed out by the field. With these figures, we already anticipate qualitatively different mechanisms for single and double ionizations.

\subsubsection{Qualitative analysis of the dynamics}

\begin{figure}
\center
        \includegraphics[width=80mm]{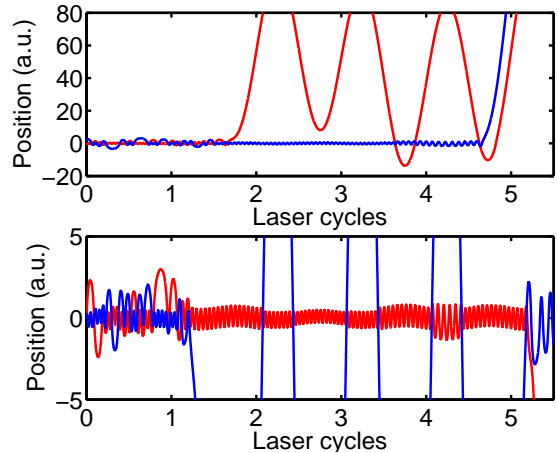}
        \caption{\label{fig:9}
        Two typical trajectories of Hamiltonian~(\ref{Hamiltonian}) for $I=10^{15}\ \mbox{W} \cdot \mbox{cm}^{-2}$ and $\omega=0.0584$ for initial conditions in the ground state energy of the helium atom. The two positions ($x$ in red and $y$ in blue) are plotted versus time (expressed in laser cycles). The recollision mechanism is seen in both panels~: In the upper one, the recollision (at the end of the panel) brings in enough energy to ionize the inner electron. In the lower panel, the recollision energy is not enough to ionize the inner electron -- the electrons exchange roles instead.
        }
\end{figure}

In this section, we describe qualitatively the dynamics based on the behavior of typical trajectories of Hamiltonian~(\ref{Hamiltonian}) in order to consider reduced dimensional models. These reduced Hamiltonians help us to explain the mechanisms occurring in phase space which lead to single, sequential double and nonsequential double ionizations. 
First we inspect individual trajectories of Hamiltonian~(\ref{Hamiltonian}). Figure~\ref{fig:9} represents two typical ionized trajectories for a given value of the laser intensity $I=10^{15} \ \mbox{W} \cdot \mbox{cm}^{-2}$ and $\omega=0.0584$. We notice that at all times the two electrons behave distincly~: While one remains close to the nucleus (the inner one) and the other one performs large excursions away from the nucleus (the outer electron). From what we have seen in the previous section, at time $t=0$, the distinction between the outer and inner electrons results from the fact that the periodic orbits $O_{x,1}$, $O_{x,2}$, $O_{y,1}$, $O_{y,2}$ organize the motion of Hamiltonian~(\ref{Hamiltonian_he}) (see Fig.~\ref{fig:6}). We notice that the role of the inner and outer electrons might be exchanged (see the lower panel of Fig.~\ref{fig:9}). 
The following scenario emerges~: The field picks up the outer electron and sweeps it away from the nucleus. This typical behavior has been observed for a large assembly of trajectories, and helps us building effective models to reveal the ionization mechanisms from a purely classical point of view. The inner electron is only driven by the interaction with the nucleus (and hence experiences nearly periodic behavior). In what follows, $y$ will always denote the inner electron and $x$ the outer one without lost of generality. For each process, we give an effective Hamiltonian obtained from Hamiltonian~(\ref{Hamiltonian}) for the inner and outer electrons.

\paragraph{Single ionization--}

Most of the single ionization appears during the ramp-up of the field (see Fig.~\ref{fig:2}).
By definition, the electron which ionizes is the outer one. Since it is far away from the nucleus, its effective Hamiltonian is
\begin{equation}
\label{seq:H1p}
{\mathcal H}_1=\frac{p_x^2}{2}+E_0 f(t)x \sin\omega t.
\end{equation}
The approximate trajectory for the outer electron is an oscillatory solution (with the period of the field) with an amplitude increasing in time~(see Figs.~\ref{fig:9} and \ref{fig:3}). Its approximate solution during the ramp-up of the field is
\begin{equation}
\label{seq:xt}
x(t)=x^0+p^0 t -\frac{E_0}{4\pi \omega^2}\left(\omega t \sin\omega t +2\cos\omega t -2 \right),
\end{equation}
starting from $x=x^0$ and $p=p^0$ at $t=0$, since $f(t)=\omega t /(4\pi)$.
If the field is not too large (during the beginning of the ramp-up of the field), the dynamics of the inner electron is governed by
\begin{equation}
\label{seq:H0}
{\mathcal H}_0=\frac{p_y^2}{2}-\frac{2}{\sqrt{y^2+1}},
\end{equation} 
which is integrable, and the inner electron is confined on a periodic orbit. Since it stays close to the nucleus, its approximate period is $T\approx 2\pi/\sqrt{2}$ obtained from the harmonic approximation. This is observed on Fig.~\ref{fig:9}.

As the outer electron moves away, the dynamics during the plateau of the laser field is described by ${\mathcal H}_1$ which is equal to
\begin{equation}
\label{seq:H1}
{\mathcal H}_1=\frac{p_x^2}{2}+xE_0 \sin\omega t,
\end{equation}
during the plateau.
This Hamiltonian has the following solutions
\begin{equation}
\label{eq:xfield}
x(t)=x^0+\left(p_x^0-\frac{E_0}{\omega} \right)t+\frac{E_0}{\omega^2}\sin\omega t,
\end{equation}
starting from $x=x^0$ and $p=p^0$ at $t=0$.
It describes, approximately, linear escape from the nucleus modulated by the action of the field (see Fig.~\ref{fig:3}).

\paragraph{Sequential double ionization--}

Once an electron has been ionized, the other electron is left with the nucleus and the field. Its effective Hamiltonian is
\begin{equation}
\label{Ham:H2}
{\mathcal H}_2=\frac{p_y^2}{2}-\frac{2}{\sqrt{y^2+1}}+yE_0\sin\omega t.
\end{equation}

\begin{figure}
\center
        \includegraphics[width=80mm]{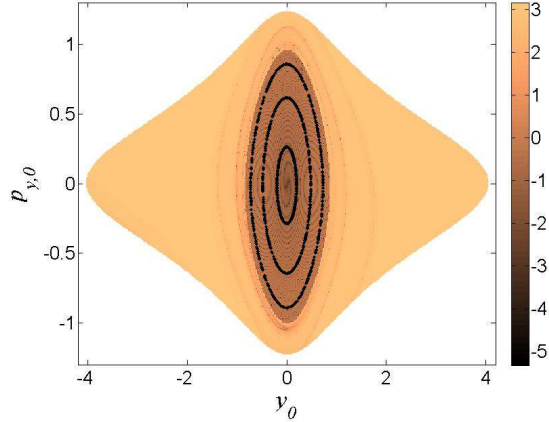}
        \caption{\label{fig:10} 
        Contour plot of $y(t)$ at time $t=215.2$~a.u. (two laser cycles) of Hamiltonian~(\ref{Ham:H2}) for $I=5\times 10^{15}\ \mbox{W} \cdot \mbox{cm}^{-2}$ and $\omega=0.0584$. Poincar\'e sections (stroboscopic plot) of selected trajectories in the elliptic central region are also depicted. The color code is on a logarithmic scale.
        }
\end{figure}

Figure~\ref{fig:10} represents a contour plot of $y(t)$ at some fixed time $t$ (two laser cycles) obtained by integrating Hamilton's equations for ${\mathcal H}_2$ from a two-dimensional space of initial conditions $(y_0,p_{y,0})$. Poincar\'e sections (stroboscopic plot with the period of the field) of some trajectories of ${\mathcal H}_2$ are also depicted in Fig.~\ref{fig:10}. This figure clearly shows two distinct regions~: The core is composed of a collection of invariant tori whose Poincar\'e sections are slight deformations of the periodic orbits obtained in the integrable case ${\mathcal H}_0$ (see Fig.~\ref{fig:4}). In this region the electron is fairly insensitive to the field (since $y$ is relatively small). The electrons in this region are the ones which do not ionize (single ionization) since the inner electron remains bounded. 
In the region outside this core, the electron is very quickly captured by the field as it is shown in Fig.~\ref{fig:10} and becomes insensitive to the soft Coulomb potential. These inner electrons ionize quickly, and their approximate trajectories are provided by the effective Hamiltonian ${\mathcal H}_1$.
The electrons outside the core region are the ones which ionize and so it corresponds to sequential double ionization.

\begin{figure}
\center
        \setlength{\unitlength}{1mm}
        \begin{picture}(80,61.5)
        \put(0,0){\includegraphics[width=80mm]{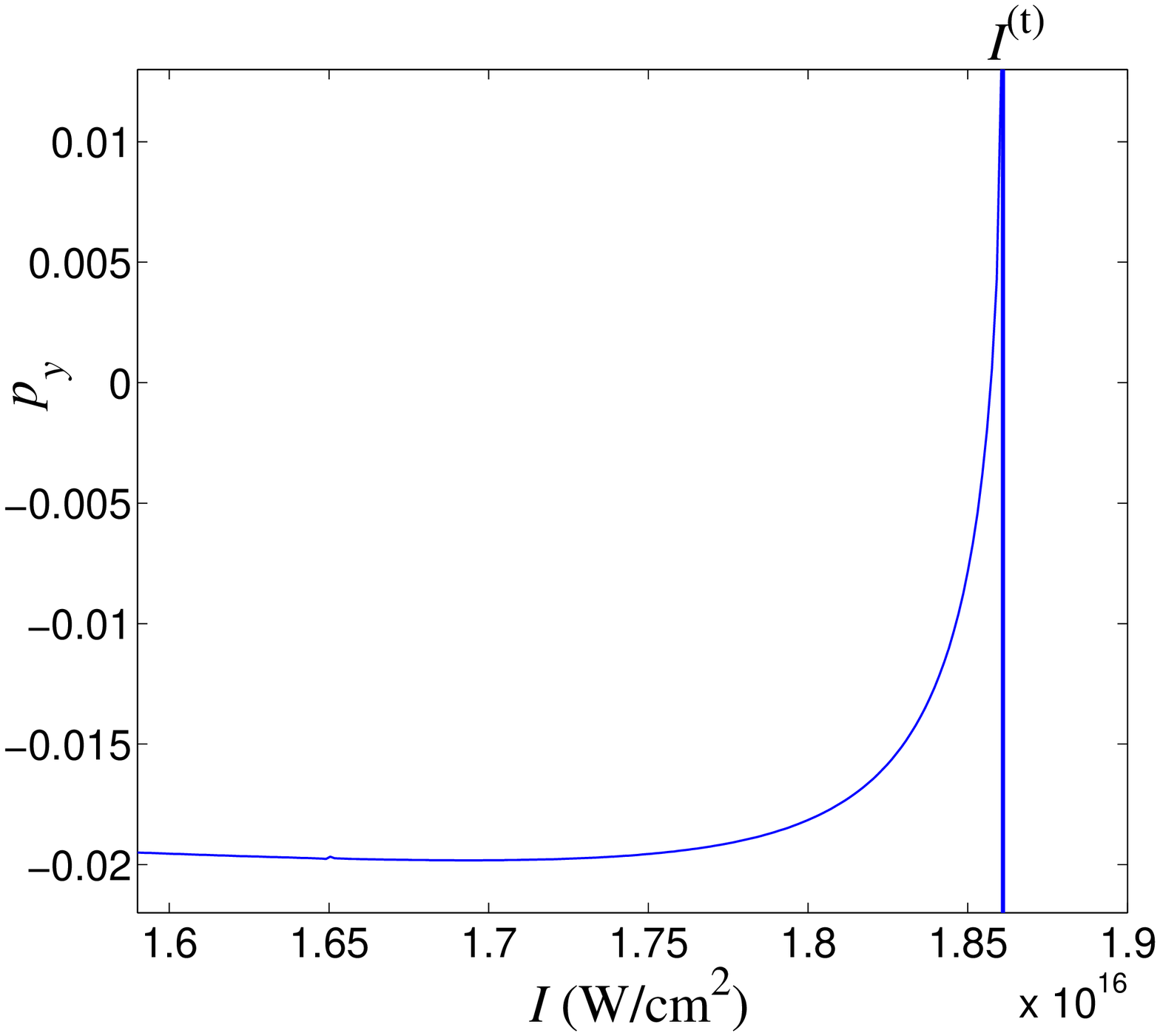}}
        \put(13,28){\includegraphics[width=40mm]{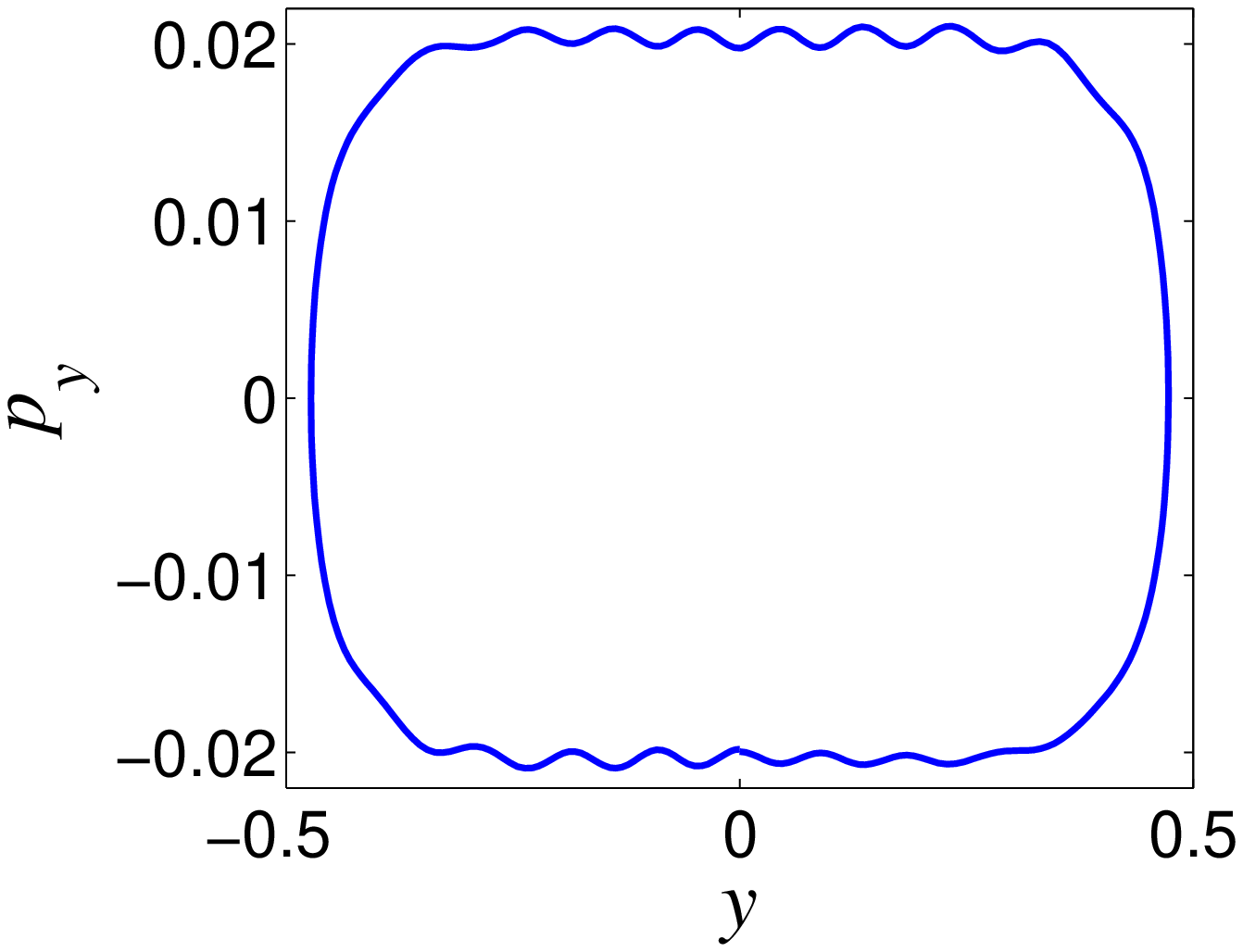}}
        \end{picture}
        \caption{\label{fig:11} 
        Momentum $p_y$ of the central periodic orbit (located at $y=0$ on the Poincar\'e section) of Hamiltonian~(\ref{Ham:H2}) for $\omega=0.0584$ as a function of the laser intensity $I$. The inset shows a projection of this periodic orbit at $I=1.7 \times 10^{16} \ \mbox{W} \cdot \mbox{cm}^{-2}$ in the $(y,p_y)$ plane.
        }
\end{figure}

The weak interaction (with the outer electron for instance) may move some particles confined on the outermost regular tori to the region where inner electrons become outer ones.
If the laser field $I$ is too small, then phase space is filled with invariant tori and no sequential double ionization occurs. The size of the elliptic region of Fig.~\ref{fig:10} (which obviously depends on $I$), is defined by a maximum escape distance from the nucleus $y_m$ (obtained when $p_y=0$). An approximation to $y_m=y_m(E_0)$ is given by the value where the potential is locally maximum~:
\begin{equation}
\label{eq:E0max}
E_0=\frac{2y_m}{(y_m^2+1)^{3/2}}.
\end{equation}
If we define $X=y_m^2+1$, this equation can be rewritten as
$$
X^3-\frac{4}{E_0^2}X+\frac{4}{E_0^2}=0,
$$
which has real solutions if 
$$
E_0\leq\frac{4}{3\sqrt{3}},
$$
which corresponds to $I\leq 2.09\times 10^{16}\ \mbox{W} \cdot \mbox{cm}^{-2}$. We notice that this derivation does not involve the laser frequency $\omega$. So we expect the intensity where the double ionization is equal to one to be approximately independent of the frequency. This rough argument can be checked using the characteristic features of the reduced dynamical model~(\ref{Ham:H2}). 
The elliptic region in Fig.~\ref{fig:10} is organized around a main elliptic periodic orbit which has the same period of the field $2\pi/\omega\approx 107.6$~a.u..
Using a Newton-Raphson algorithm, we have computed the location and the stability of this periodic orbit for $\omega=0.0584$. As long as this periodic orbit exists, it remains elliptic in the whole range of intensities we have considered. On the stroboscopic plot (with frequency $\omega$) the periodic orbit is located at $y=0$. In Fig.~\ref{fig:11}, the coordinate $p_y$ of the periodic orbit on the stroboscopic plot is represented as a function of $I$. We notice that for a large set of intensities this periodic orbit is located close to the center at $p_y\approx 0$.
For intensities larger than $I^{(t)}=1.86 \times 10^{16}\ \mbox{W} \cdot \mbox{cm}^{-2}$, the periodic orbit does not exist, and no major island of regularity remain. Therefore, it is expected that the sequential double ionization probability is equal to one in this range of intensities, as observed on the probability curve on Fig.~\ref{fig:1}. We notice that $I^{(t)}$ is close to the value obtained using a rough estimate of $y_m$. 

We argue that this (short) elliptic periodic orbit of the reduced model~(\ref{Ham:H2}) is central to the understanding of the amount of sequential double ionization probability. A natural question concerns the correspondence of this periodic orbit in the full Hamiltonian model~(\ref{Hamiltonian}). Of course, it cannot be a periodic orbit in the full model, since a trajectory initiated on a periodic orbit cannot ionize. In fact, it corresponds to a structure which is unbounded (but mostly regular) in the reduced phase space of one electron, and a periodic orbit (hence bounded) in the reduced phase space of the other electron.

\paragraph{Nonsequential double ionization}

As we have seen before, when the field is turned on, its action is concentrated on only one electron-- the outer one -- as a first step. The field drives the outer electron further from the nucleus, leaving the other electron nearly unaffected by the field because the amplitude of the field is proportional to time $t$. In the pulse plateau, the outer electron far from the nucleus might come back close to the nucleus if the field strength is not too large [see Eq.~(\ref{eq:xfield})]. Then it can transfer part of its energy to the inner electron. This is the recollision scenario~\cite{cork93,scha93} in purely classical terms (i.e.\ without tunneling). 

From then on, two outcomes are possible~: If the energy brought back by the outer electron is sufficient for the inner  electron to escape from the regular region, then it might ionize together with the outer electron.
The maximum energy ${\mathcal E}_x$ of the outer electron when it returns to the inner region (after having left the inner region with a small momentum $p_0$ close to zero) is obtained from Hamiltonian~(\ref{seq:H1p}) and is equal to ${\mathcal E}_x=\kappa U_p$ where $U_p=E^2_0/(4\omega^2)$ is the ponderomotive energy and $\kappa=3.17314...$ is the maximum recollision kinetic energy in units of $U_p$~\cite{cork93,beck94,band05}. Here we complement the recollision scenario (which focuses on the outer electron) by providing the phase space picture of the inner electron~: In order to ionize the inner electron, the energy brought back by the outer electron has to be of order of the energy difference between the core ($y=0$) and the boundary ($y=y_m$) of ${\mathcal H}_2$ (see Fig.~\ref{fig:10}) which is equal to 
\begin{equation}
\label{seq:DEy}
\Delta {\mathcal E}_y=2-\frac{2}{\sqrt{y_m^2+1}}.
\end{equation}
The equal-sharing relation which links the classical picture of the outer electron $x$ with the one of the inner electron $y$, 
\begin{equation}
\label{eq:DEyEx}\Delta {\mathcal E}_y=\frac{{\mathcal E}_x}{2},\end{equation}
defines (through an implicit equation) the expected value of the field $E_0^{(c)}$ where the maximum nonsequential double ionization occurs, because it describes the case where each outer electron acquires enough energy from the field to potentially ionize the inner electron, while remaining ionized itself. The equation which links $E_0^{(c)}$ to $y_m$ is given by Eq.~(\ref{eq:E0max}).
In order to solve Eq.~(\ref{eq:DEyEx}), we define $\eta=1/\sqrt{y_m^2+1}$ and this equation becomes~:
$$
\eta^4(1+\eta)=\eta_0^4,
$$
where $\eta_0^2=2\omega/\sqrt{\kappa}$.
For small $\omega$, the expansion of $E_0^{(c)}$ is given by $E_0^{(c)}=2\eta^2-\eta^4+O(\eta^6)$ where $\eta$ is solution of the equation above and it leads to
\begin{equation}
\label{seq:Ec}
E_0^{(c)}=\frac{4\omega}{\sqrt{\kappa}}-\left(\frac{2\omega}{\sqrt{\kappa}}\right)^{3/2}-\frac{7}{32}\left(\frac{2\omega}{\sqrt{\kappa}}\right)^{5/2}+O\left(\left(\frac{2\omega}{\sqrt{\kappa}}\right)^{3} \right).
\end{equation}

\begin{figure}
\center
        \includegraphics[width=80mm]{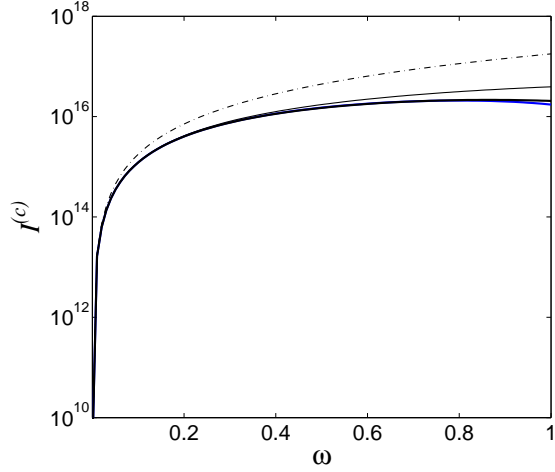}
        \caption{\label{fig:12} 
        Intensity $I^{(c)}$ as given by Eq.~(\ref{seq:Ec}) where the nonsequential double ionization is expected to be maximum. The blue curve is determined numerically by solving Eq.~(\ref{eq:DEyEx}). The black bold curve is the approximation given by Eq.~(\ref{seq:Ec}). The thin black curve is obtained by dropping the term $O((2\omega/\sqrt{\kappa})^{5/2})$ in Eq.~(\ref{eq:DEyEx}), and the thin dashed dotted curve is obtained when the term $O((2\omega/\sqrt{\kappa})^{3/2})$ is neglected.}
\end{figure}

\begin{figure}
\center
        \includegraphics[width=80mm]{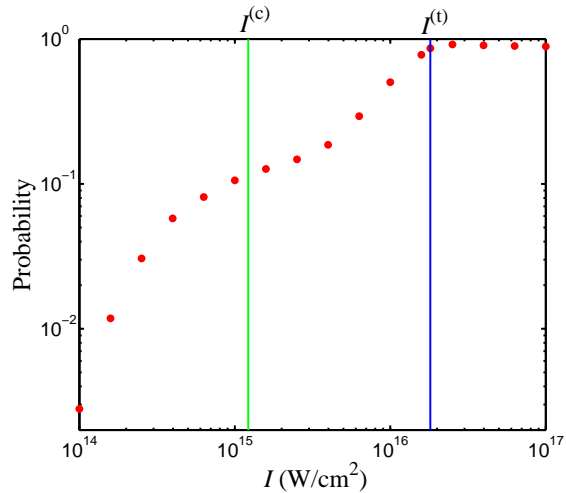}
        \caption{\label{fig:13} 
        Double ionization probability for Hamiltonian~(\ref{Hamiltonian}) for $\omega=0.1$ as a function of the intensity of the field $I$. We use a microcanonical initial set of~$3000$ trajectories randomly chosen in the accessible phase space of the helium atom in the ground state of energy. The field is a 2-4-2 laser cycle pulse shape (see Fig.~\ref{fig:3}).
        The vertical lines indicate (in green) the laser intensity $I^{(c)}\approx 1.22 \times 10^{15}\ \mbox{W} \cdot \mbox{cm}^{-2}$ where our dynamical analysis predicts the maximum of nonsequential double ionization, and (in blue) the intensity $I^{(t)}\approx 1.81 \times 10^{16}\ \mbox{W} \cdot \mbox{cm}^{-2}$ where the double ionization is expected to be complete.}
\end{figure}

Figure~\ref{fig:12} represents $I^{(c)}$ as a function of $\omega$ as given by the various truncations of the above formula. It shows that the truncation to the first three terms in Eq.~(\ref{seq:Ec}) is accurate for a wide range of laser frequencies. To leading order the corresponding intensity varies as $\omega^2$. However, higher order terms in the expansion in $\omega$ are necessary to obtain a quantitative agreement. For $\omega=0.0584$, the approximate intensity given by Eq.~(\ref{seq:Ec}) is $4.58\times 10^{14}\ \mbox{W} \cdot \mbox{cm}^{-2}$ which is in excellent agreement with $I^{(c)}$ (see Fig.~\ref{fig:1}). We have checked that this relation between $E_0^{(c)}$ and $\omega$ holds for a wide range of values of the laser frequency, i.e.\ for $\omega$ between 0.01 and 1. 
We have plotted the double ionization probability as a function of the intensity in Fig.~\ref{fig:13} for another value of the laser frequency $\omega=0.1$. The intensity $I^{(c)}\approx 1.22\times 10^{15} \ \mbox{W} \cdot \mbox{cm}^{-2}$ correctly locates the maximum of the nonsequential double ionization probability and hence the ``knee''. To complete the double ionization picture at $\omega=0.1$, a computation of the location of the central elliptic periodic orbit as performed in Fig.~\ref{fig:11} predicts that at $I^{(t)}\approx 1.81\times 10^{16}\ \mbox{W} \cdot \mbox{cm}^{-2}$, this periodic orbit disappears and can no longer organize the regular motion in the inner region. Therefore at this value, a probability close to one is expected for SDI, in agreement with Fig.~\ref{fig:13}. We notice that the value for $I^{(t)}$ is close to the one obtained for $\omega=0.0584$ as expected from the approximate independence of $I^{(t)}$ with respect to $\omega$.

In summary, when the field $I$ is too small, then the outer electron cannot gain enough energy to ionize the innermost electron. So only the inner electron on the boundary of the regular region where there are fewer electrons as the intensity is decreased, do participate in nonsequential double ionization. When the field $I$ is too large (typically of order $I^{(t)}$ or larger), there are too few inner electrons to contribute to nonsequential double ionization since the size of the regular region (as given by $y_m$) vanishes and the volume of phase space housing the inner electron  becomes too small. These two considerations explain the (expected asymmetric) bell shape of the resulting nonsequential double ionization probability, which, when put together with the monotonic rise of the SDI at higher intensities, adds up to the knee in question.

\section{Conclusion}
In this article, we have considered the classical dynamics of sequential and nonsequential double ionization in the helium atom. In the absence of the field, the dynamics shows highly chaotic dynamics without any regular regions. Under the influence of a strong laser field, this atom undergoes single, sequential double and nonsequential double ionizations. The ``knee'' of the double ionization probability as a function of the laser intensity, observed in experiments and numerical simulations, is the sum of two probabilities, each one associated with a different process. Their sum leads to the knee shape.
Using qualitative and quantitative analysis of the dynamics, we provide a physical mechanism for each of these processes. Each argument supporting these mechanisms comes from low-dimensional and even integrable Hamiltonian systems, and are based on the effective separation between an inner and outer electron. This separation results from the four periodic orbits which organize the chaotic dynamics of the helium atom without the laser field. The nonsequential double ionization results from a transfer of energy from the outer electron to the inner electron (i.e.\ a recollision between these two). An approximate model based on low-dimensional effective Hamiltonians predicts the value at which its probability is maximum. The sequential double ionization results more directly from interaction between the field with the two electrons individually. Based on this scenario, a dynamical analysis in terms of periodic orbits is able to predict accurately the value at which the double ionization probability is expected to be equal to one. 

\ack

CC acknowledges financial support from the PICS program of the CNRS. This work is partially funded by NSF. We thank A. D. Bandrauk, W. Becker, and S. L. Haan for useful discussions.



\end{document}